%% file: fpga_transformer_v2.tex
\documentclass{article}

\usepackage[final,nonatbib]{amlc_guide}

\usepackage[utf8]{inputenc}
\usepackage[T1]{fontenc}
\usepackage{hyperref}
\usepackage{url}
\usepackage{booktabs}
\usepackage{amsfonts}
\usepackage{amsmath}
\usepackage{nicefrac}
\usepackage{microtype}
\usepackage{xcolor}
\usepackage{graphicx}
\usepackage{multirow}
\usepackage{placeins}
\usepackage{tikz}
\usetikzlibrary{shapes.geometric, arrows.meta, positioning, fit, calc, backgrounds}
\usepackage[font=small,labelfont=bf]{caption}

\title{Enabling AWS Cloud FPGA Development Using High-Level Vitis HLS}

\author{%
  David Rostcheck \\
  \texttt{david@rostcheck.com} \\
}

\begin{document}

\maketitle

\begin{abstract}
We present a toolchain for development of ML models on AWS EC2 F2 cloud FPGA infrastructure from high-level language, implementing a complete Vitis HLS $\to$ HDK bridge $\to$ AFI workflow that has been absent from the academic literature despite eight years of AWS FPGA availability. We introduce a four-level methodology (FP32 PyTorch $\to$ bit-accurate quantized Python $\to$ HLS C-simulation $\to$ physical FPGA) to aid in localizing errors, including FPGA-specific circuit errors. Together these tools allow coding agents to effectively implement FPGA designs from user directions, testing and debugging as needed. We document the deployment workflow, timing closure strategies, and practical HLS engineering decisions that proved crucial. We validate using TinyBERT fine-tuned for address Named Entity Recognition, implementing the embedding layer and the first of four encoder layers on the FPGA hardware. Our implementation streams weights from host DRAM, yielding poor performance compared to CPU/GPU implementations; we discuss forward paths.
\end{abstract}

\section{Introduction}
\label{sec:introduction}

Engineering demanding workloads such as ML inference requires coordinated optimization between the hardware architecture and software operations. The transformer-based software stack that represents current state of the art in ML implements a neural network via matrix multiplication operations. These operations are accelerated by high-performance hardware GPUs with on-board memory. Transformer inference is increasingly constrained by memory bandwidth rather than compute~\cite{vaswani2017attention,dettmers2022gptint8,ivanov2021data,pope2023efficiently}.

In comparison with the essentially von Neumann-based CPU/GPU architecture, Field Programmable Gate Array (FPGA) devices represent an alternative hardware platform. FPGAs offer memory distributed throughout the device and feature programmable circuitry, allowing memory and compute to be placed together in a more distributed way. These features support deterministic latency and fixed power draw independent of request rate, and could permit more efficient designs, particularly for implementing smaller models and those where GPU tail-latency variance and idle power are concerns~\cite{yan2024fpga_ml_survey,flightllm}.

Research in generative AI using FPGA devices has been constrained, however, by the difficulty of programming ML systems on them. Despite eight years of availability (F1 since 2017, F2 since December 2024~\cite{aws_f2_blog}), AWS cloud FPGA instances have not been used for ML inference in any published academic work, although their pay-on-demand cost model lowers the funding barrier for research projects. The University of Toronto noted in 2024 that F1 instances ``have not been used for such large ML applications''~\cite{gao2024multifpga}. The gap is not hardware capability; the AMD VU47P on F2 has 9,024 DSP slices, 16\,GiB HBM, and 64\,GiB DDR4. We identify the absence of a documented, reproducible workflow from HLS C++ kernels to deployed Amazon FPGA Images (AFIs) as the barrier.

This paper intends to close this gap. We present a complete toolchain and methodology for developing ML models on AWS EC2 F2 using the complete Vitis HLS (high level language) $\to$ HDK bridge $\to$ AFI workflow for ML workloads. Supporting contributions include a four-level validation methodology (\S\ref{sec:validation}), build time characterization for the AWS F2 HDK workflow~\cite{aws_f2_hdk} (\S\ref{sec:deployment}), and a timing closure analysis for SSI FPGA devices under ML workloads (\S\ref{sec:timing}).

\section{System design}
\label{sec:design}
Our primary research goal was to characterize a working toolchain and analysis path by which a transformer could be implemented on an AWS F2 FPGA. This goal led to the following design decisions (detailed more fully in following subsections). Taking true advantage of the FPGA's architectural capabilities would require different choices, discussed in Section~\ref{sec:future_work}:
\begin{enumerate}
    \item \textbf{High level language:} For ease of implementation we chose the C++-based Vitis HLS language instead of lower-level Verilog HDL.
    \item \textbf{Community bridge:} The F2 toolchain does not natively support creating an AFI image from high-level Vitis HLS. We used a community-based open source bridge~\cite{boonhong2024bridge}  to facilitate this path.
    \item \textbf{W16A16 quantization:} To better debug numerical issues resulting from FPGA design logic, we chose INT16 weights and INT16 activations (vs. more common INT8 quantizations)
    \item \textbf{Four-level validation methodology:} To localize where in the stack errors were appearing, we developed a four-level validation methodology described further in Section~\ref{sec:validation}
\end{enumerate}

\subsection{FPGA architecture}
\label{sec:fpga_architecture}
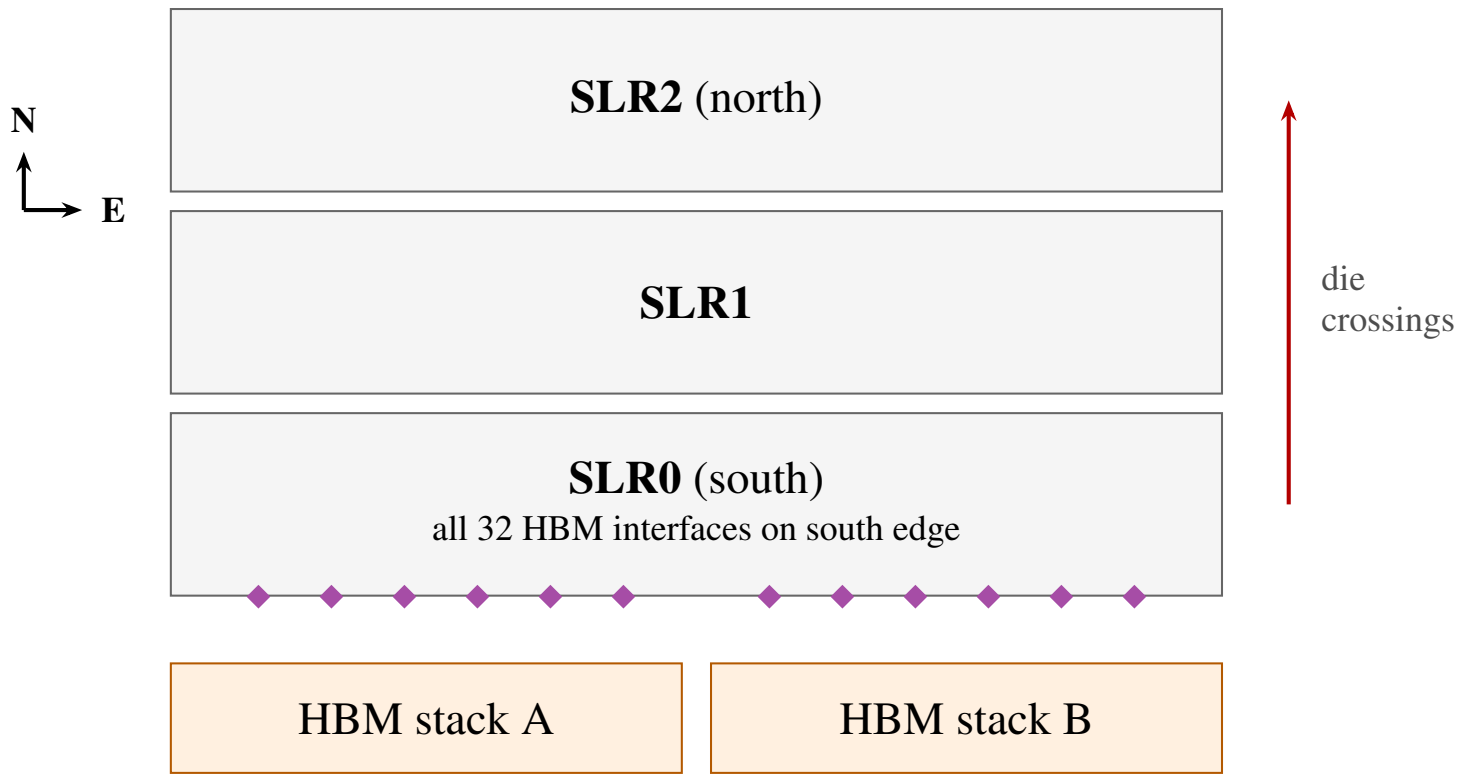
\begin{figure}
\centering
\resizebox{0.7\columnwidth}{!}{%
\begin{tikzpicture}[
    slr/.style={rectangle, draw=black!60, fill=gray!8, minimum width=7.2cm,
        minimum height=1.25cm, align=center, font=\footnotesize},
    hbm/.style={rectangle, draw=orange!70!black, fill=orange!12, minimum width=3.5cm,
        minimum height=0.75cm, align=center, font=\footnotesize},
    hbmif/.style={diamond, fill=violet!70, inner sep=1.2pt},
    cross/.style={-{Stealth[length=4pt]}, thick, red!70!black},
    lbl/.style={font=\scriptsize\color{black!70}},
]

\node[slr] (slr2) {\textbf{SLR2} (north)};
\node[slr, below=0.12cm of slr2] (slr1) {\textbf{SLR1}};
\node[slr, below=0.12cm of slr1] (slr0) {\textbf{SLR0} (south)\\[-1pt]\scriptsize all 32 HBM interfaces on south edge};

\node[hbm, below=0.45cm of slr0.south west, anchor=north west] (hbma) {HBM stack A};
\node[hbm, below=0.45cm of slr0.south east, anchor=north east] (hbmb) {HBM stack B};

\foreach \x in {-3.0,-2.5,-2.0,-1.5,-1.0,-0.5,0.5,1.0,1.5,2.0,2.5,3.0}
    \node[hbmif] at ($(slr0.south) + (\x,0)$) {};

\draw[-{Stealth[length=4pt]}, thick] ([xshift=-1.0cm]slr1.north west) -- ++(0,0.4) node[above, font=\scriptsize\bfseries] {N};
\draw[-{Stealth[length=4pt]}, thick] ([xshift=-1.0cm]slr1.north west) -- ++(0.4,0) node[right, font=\scriptsize\bfseries] {E};

\draw[cross] ([xshift=0.45cm]slr0.east) -- ([xshift=0.45cm]slr2.east);
\node[lbl, anchor=west, align=left] at ([xshift=0.55cm]slr1.east) {die\\crossings};

\end{tikzpicture}%
}
\caption{Top-down view of the AMD VU47P. The three SLRs are separate dies laid side by side on a shared silicon interposer, drawn here from south to north. The two HBM stacks sit on the same interposer, south of SLR0, and all 32 HBM interface sites (violet diamonds, not to scale) are on SLR0's south edge.}
\label{fig:floorplan}
\end{figure}

While FPGAs provide programmable hardware, the physical architecture of the device remains important for combined hardware/software optimization. The AWS F2 instance hosts an AMD VU47P Virtex UltraScale+ FPGA which is organized in three separate dies called Super-Logic Regions (SLRs), which are  co-located and wired together (see Figure~\ref{fig:floorplan}). Crossing an SLR boundary requires traversing a long wire, which may cause signal timing problems.

Memory in an FPGA is distributed throughout SLRs and is available in a hierarchy of layers that are progressively both larger and farther away from individual logic elements, as shown in Table~\ref{tab:memory_hierarchy}.

\begin{table}
\caption{Memory hierarchy of the AMD VU47P on AWS F2, ordered from closest to compute to farthest. LUTRAM is carved from the same LUTs used for logic, so its totals are an upper bound. Latency is in clock cycles for on-chip memory; off-chip memory is characterized by peak bandwidth. Near-chip High Bandwidth Memory is a major feature of the VU47P but the community bridge needed to produce an Amazon FPGA Image (AFI)  from high level HLS code did not provide access to it, a major limitation.}
\label{tab:memory_hierarchy}
\centering
\small
\begin{tabular}{@{}lrrrl@{}}
\toprule
Memory type & Element size & Elements & Total & Latency / bandwidth \\
\midrule
Flip-flops & 1\,bit & 2,607,360 & 2.6\,Mb & 1 cycle \\
LUTRAM & 64\,bits & up to ${\sim}600{,}000$ & up to 36.7\,Mb & ${\sim}1$ cycle \\
BRAM & 36\,Kb & 2,016 & 70.9\,Mb & 1--2 cycles \\
URAM & 288\,Kb & 960 & 270\,Mb & multi-cycle \\
\midrule
HBM & 8\,GiB & 2 & 16\,GiB & up to 460\,GB/s \\
DDR4 & 64\,GiB & 1 & 64\,GiB & ${\sim}25$\,GB/s \\
\bottomrule
\end{tabular}
\end{table}

\subsection{Toolchain architecture}
\label{sec:toolchain}
FPGA devices are natively programmed in a low-level Hardware Description Language (HDL), the most popular being Verilog/SystemVerilog and VHDL. When fed into a synthesis tool ("Vivado" on the AMD/Xilinx FPGAs that the AWS F2 uses) the HDL code produces a physical design that specifies the direct placement of logic cells, lookup tables (LUTs), and routing, to which the FPGA is then configured. This level of implementation is termed Register-Transfer Level (RTL) logic. While conceptually comparable to an assembler producing machine code, the physical design is literally a circuit design which the FPGA directly manifests via a grid (array) of programmable logic gates, as compared with the set of CPU instructions that machine code represents. The physical design can thus, for example, specify many paths that execute in parallel by physically manifesting different circuit paths. The Vivado synthesis tool performs more sophisticated operations than an assembler, making use of simulation to optimize the timing and physical placement of the circuit design.

While HDL provides the most control, its verbose low-level nature can prove a barrier to quickly implementing interesting ML designs. We chose to target a High Level Synthesis (HLS) language, the C/C++-based "Vitis HLS", which the Vitis tool synthesizes to lower-level HDL.

The Hardware Development Kit (HDK) for the AWS F2~\cite{aws_f2_hdk} provides a shell that separates AWS-managed infrastructure (PCIe, DDR controllers, DMA) from user custom logic (CL). The shell handles infrastructure functions and provides defined interface points to which the user-supplied CL integrates. Unlike the user/kernel separation in a traditional OS, the shell and user CL are combined by the Vivado synthesizer to produce the RTL/circuit design, so both shell and user CL ultimately directly manifest hardware. Conceptually the shell is more similar to the set of linked libraries that operate the loop and pseudo-OS functions in a Real-Time Development Kit (RTK).

Pre-created CL for common tasks can also be imported using "IP cores" which play a role analogous to reusable libraries in CPU-based programming. 

The HDK Custom Logic (CL) flow creates an AFI file, which is loaded to the FPGA. The higher-level Vitis HLS that we target cannot produce an AFI directly. To bridge this gap, we used a community-developed bridge to allow end-to-end flow from Vitis HLS to an AFI, as shown in Figure~\ref{fig:flow}.

\begin{figure}[!htb]
\centering
\fbox{\parbox{0.94\columnwidth}{\centering\small
Vitis HLS C++ $\to$ RTL export $\to$ \texttt{hls\_to\_cl.py} (bridge) $\to$ HDK CL project $\to$ post-bridge patches $\to$ DCP build (Vivado) $\to$ AFI creation $\to$ physical FPGA}}
\caption{Toolchain from Vitis HLS C++ to a deployed F2 instance. The community bridge (\texttt{hls\_to\_cl.py}) and the post-bridge patches are the only steps that are neither our code nor standard AWS or AMD tooling; they are the source of the limitations in Section~\ref{sec:bridge_limitations}, including the single DDR port and lack of HBM access.}
\label{fig:flow}
\end{figure}

\subsection{Bridge limitations}
\label{sec:bridge_limitations}
While it did allow for creation of a loadable AFI from Vitis HLS code, the community-supported bridge imposed several limitations, which caused later downstream issues:
\begin{itemize}
    \item \textbf{Single memory master.} The bridge supports exactly one \texttt{m\_axi} port and one \texttt{s\_axilite} control interface per kernel\footnote{AXI: Advanced eXtensible Interface, a bus standard for moving data between hardware blocks.}. Having only a single memory access port pair significantly limits the design's memory access bandwidth, since higher bandwidth would require multiple channels with multiple AXI ports.
    \item \textbf{DDR only; no HBM.} The bridge template connects the single \texttt{m\_axi} port directly to the DDR controller (memory on the EC2 host), not to the FPGA's near-chip High Bandwidth Memory (HBM), a major limitation.
    \item \textbf{Fixed 512-bit data width.} The template assumes a 512-bit AXI data path and does not adapt to the data width of the generated kernel port.
    \item \textbf{No multi-die (SSI) awareness.} The bridge does not insert SLR-crossing pipeline registers or provide SLR placement constraints.
    \item \textbf{Incomplete IP packaging.} The bridge does not copy the VHDL sub-IP files for HLS-generated floating-point cores (\texttt{fdiv\_32}, \texttt{fpext}) into the CL project.
\end{itemize}

\subsection{Model and task}
\label{sec:model_and_task}
For a testbed model we chose TinyBERT~\cite{jiao2020tinybert}, a 4-layer, 312-hidden-dimension distilled BERT variant~\cite{hinton2015distilling,sanh2019distilbert} with $\sim$14.5M parameters, primarily because it was small and well understood. We fine-tuned the \texttt{huawei-noah/TinyBERT\_General\_4L\_312D} pretrained checkpoint using HuggingFace Transformers~\cite{wolf2020transformers} with AdamW, learning rate $5 \times 10^{-5}$, 5 epochs.

For a test task we chose name-entity recognition (NER). We fine-tuned on a synthetically generated BIO-tagged address NER dataset (560 train / 120 val / 500 test samples). The dataset generator (\texttt{data/generate\_address\_ner\_dataset.py}) is released with the code. Entity types are STREET, CITY, STATE, ZIP, and PERSON.

\subsection{System architecture}
\label{sec:system_architecture}
Figure~\ref{fig:architecture} shows the system architecture. The host loads tokenized input and packed W16A16 weights into DDR4 via PCIe. The CL kernel processes the four-layer TinyBERT pipeline through a single \texttt{m\_axi} DDR port: each stage reads its weight matrix and input activations, computes, and writes output activations back to DDR at fixed offsets. This sequential DDR round-trip is the primary latency bottleneck in the current design.

\input{figure_architecture}

\paragraph{Memory architecture:} We originally stored the weights in BRAM. Our first design resulted in a deep LUT mux tree that could not close 250 MHz timing. A redesign partitioned weights in BRAM into banks, but the transformer matmul (a fundamentally global operation) created cross-SLR routing congestion that also could not close timing. HBM would have been ideal for this case, but since the bridge can not address it (Section~\ref{sec:bridge_limitations}), we needed to store weights in DRAM. Still, BRAM working storage spanned two of the three SLRs (SLR0 53.5\%, SLR1 46.9\%, with SLR2 at 3.9\% holding only the shell's DDR controller), forcing the working footprint across an SLR boundary and continually causing timing violations. The issue was finally resolved by moving the largest ${\sim}$600k array to URAM, clearing timing (WNS = +0.027ns).

\subsection{W16A16 quantization architecture}
\label{sec:quantization}
\paragraph{Why W16A16:}
We began with W8A8 (INT8 weights and activations with per-channel/per-token dynamic scaling). Two problems emerged. First, INT8 per-token quantization produces large intermediate errors (maximum 86.0 at the FFN stage for a known outlier activation of ${\sim}4528$ in FP32). While these errors often cancel at the argmax level, they make layer-by-layer FPGA verification intractable: when each stage must be numerically compared against the reference, errors of this magnitude obscure the distinction between quantization noise and implementation bugs. Second, W8A8 generated timing violations on the FPGA: the per-token quantization division produced a hardware divider that failed timing ($\text{WNS} = -1.148$\,ns), and FP32 intermediate operations could not meet the 250\,MHz clock constraint. W16A16 resolves both issues: the division is replaced by a reciprocal multiply (Section~\ref{sec:quantization}) and maximum intermediate error drops to ${\leq}\,0.137$ across all encoder stages, enabling the four-level validation methodology (Section~\ref{sec:validation}). 

Better quantization schemes for optimizing memory and model size are available, particularly those described in~\cite{bondarenko2021understanding}. However, FPGA development is, at its core, electrical engineering, not computer programming. As such, it imposes significant and fundamental differences on error analysis. In traditional CPU/GPU-based programming, mathematical operations themselves can be assumed to work. This is not the case in FPGA design synthesis, where these operations represent circuits that can be malformed, or can use IP cores that do not work properly in the accuracy range requested. Using W16A16 with generous accumulators, while not optimal for efficiency, allowed sufficient headroom to identify when errors were being introduced so they could be localized using the four-level validation method described in \S\ref{sec:validation}.

\paragraph{Quantization:} We used INT16 weights $\times$ INT16 activations (W16A16) with \texttt{ap\_fixed<32,16>} element-wise operations and \texttt{ap\_fixed<64,32>} dequantization. Each weight matrix row is quantized to INT16 using per-channel absmax scaling:
\begin{equation}
w_q = \text{round}\!\left(\frac{w}{\text{absmax}}\right) \times 32767
\end{equation}
The absmax value (not absmax/32767) is stored as the per-channel scale in \texttt{ap\_fixed<32,16>}.

Each token's activation vector is quantized to INT16 using per-token absmax. The division is performed as a float32 reciprocal followed by a per-element multiply (see Non-linear operations below).

INT16 $\times$ INT16 products accumulate in \texttt{ap\_int<48>} (sufficient for 312-element dot products without overflow). The four-step wide dequantization, all in \texttt{ap\_fixed<64,32>}, is:
\begin{equation}
\text{output} = \text{acc} \times w_{\text{scale}} \times a_{\text{absmax}} \times \frac{1}{32767} + \text{bias}
\end{equation}

\paragraph{Non-linear operations:} Our initial implementation used \texttt{hls\_math.h} vendor functions throughout. In practice, we found that \texttt{hls::exp()} on \texttt{ap\_fixed<32,16>} returns zero on physical FPGA after place-and-route (an apparent bug in the HLS-generated RTL core); \texttt{hls::erf()} promotes to FP64, generating \texttt{dmul\_64}/\texttt{dadd\_64} IP cores that violate 250\,MHz timing; and 64-bit fixed-point division generates an \texttt{sdiv\_80ns\_48ns\_80} hardware divider (WNS~$= -1.148$\,ns). We replaced each with a targeted alternative: (1)~softmax \texttt{exp} uses a 2048-entry LUT covering $[-11, 0]$; (2)~GELU uses a 4096-entry LUT with linear interpolation covering $[-8, 8]$ (max error $< 3 \times 10^{-5}$ vs.\ FP64, 30$\times$ better than a tanh approximation); (3)~activation quantization division uses a float32 reciprocal (\texttt{inv\_am = 1.0f / (float)am}) followed by per-element multiply. A fourth issue, dequantization overflow for large accumulator products (e.g., FFN dim~3 outlier: $\sim$69B, overflowing \texttt{ap\_fixed<64,32>}'s $\pm$2.1B range), was resolved by widening steps~1--2 of the four-step dequantization to \texttt{ap\_fixed<80,48>} before truncating. As a timing closure technique, the GELU output array ($128 \times 1200 = 600$\,KB BRAM) was eliminated by computing GELU twice (once for absmax, once for quantization), avoiding an SLR-crossing BRAM placement that added 2.3\,ns logic + 2.7\,ns routing. Fuketa et~al.~\cite{fuketa2025} take a contrasting approach, replacing all transcendentals with polynomial matrix approximations; our hybrid LUT strategy prioritizes accuracy (vendor-optimized or validated LUT) while avoiding the timing hazards of large IP cores.

\subsection{HLS kernel design}
\label{sec:kernel}

The kernel (\texttt{tinybert\_ner\_axi}) is a single HLS C++ function with one \texttt{m\_axi} DDR port (\texttt{ap\_uint<512>*}) and one \texttt{s\_axilite} control interface. All weights, activations, and debug outputs share a single DDR buffer (required by the bridge PoC's single-port constraint).

\paragraph{Pipeline:} Token IDs $\to$ Embedding (lookup + LayerNorm) $\to$ 4$\times$ Encoder Layer (QKV projection + scaled dot-product attention + output projection + LayerNorm + FFN + LayerNorm) $\to$ Classification Head $\to$ Logits.

\paragraph{Streaming matmul:} Weight matrices are streamed from DDR in tiles rather than stored on-chip. This eliminates BRAM cascade timing violations (WNS = $-1.399$\,ns with on-chip weights) at the cost of DDR bandwidth and inference latency, which is dominated by sequential DDR weight reads rather than compute. The 512-bit AXI bus width requires 64-byte-aligned writes; a read-modify-write pattern creates a 6.5\,ns critical path that violates the 4.0\,ns budget. Table~\ref{tab:resources} gives the resulting device utilization for the embedding layer plus encoder layer~0 build.

\begin{table}[t]
\caption{Device-total resource utilization on AMD VU47P (AWS F2), post-route for the
embedding + encoder-layer-0 build (includes the AWS shell). Logic (LUT/FF/DSP) is $<$6\%, but
\emph{Block RAM is the binding resource at 36\% of tiles}, almost all of it
custom logic (CL); the AWS shell contributes only ${\sim}$27 of the 727.5 BRAM tiles. BRAM is reused across encoder
layers, so it does not scale with the full 4-layer pipeline.}
\label{tab:resources}
\centering
\small
\begin{tabular}{@{}lrrr@{}}
\toprule
Resource & Used & Available & Utilization \\
\midrule
CLB LUT         & 73,272 & 1,301,552 & 5.6\% \\
CLB FF          & 90,025 & 2,603,104 & 3.5\% \\
Block RAM tile  & 727.5  & 2,016     & \textbf{36.1\%} \\
DSP             & 138    & 9,024     & 1.5\% \\
URAM            & 40     & 960       & 4.2\% \\
\bottomrule
\end{tabular}

\vspace{2pt}
{\footnotesize Per-SLR BRAM-tile occupancy (shell+CL): SLR0 53.5\%, SLR1 46.9\%,
SLR2 3.9\%. The custom-logic BRAM occupies the two lower SLRs (SLR0 just over
half-full), while SLR2 holds only the shell's DDR controller. This near-even
split across the SLR0--SLR1 boundary constrains placement of the memory-heavy
attention datapath.}
\end{table}

\subsection{Four-level validation methodology}
\label{sec:validation}

We validate correctness through four implementation levels (Figure~\ref{fig:validation}). The L2 intermediate is the key contribution: a Python reference that replicates HLS \texttt{ap\_fixed} arithmetic exactly, enabling per-stage error isolation without FPGA builds. When L3 or L4 output diverges from L1, L2 isolates whether the cause is quantization design (L1 vs L2 mismatch) or implementation error (L2 vs L3/L4 mismatch). For example, when implementing the first TinyBERT layer we observed incorrect results in L4, while L3 was correct: the logic and code were correct, but there was an error specific to the FPGA implementation. Investigation via targeted builds eventually determined that the HLS-to-CL bridge had not copied the .dat initialization file into the synthesis directory. Vivado silently zero-filled the ROM, so on hardware the LUT returned zero for every input. The 4-level validation serves as an effective loop closure tool, allowing the coding agent to autonomously localize and diagnose  problems.

\input{figure1_validation}

\subsection{AI agent use and research methodology}
\label{sec:ai_use}

We conducted the work during a 1-month research spike, which included initial paper writeup. The team consisted of a single human researcher using an AWS Kiro IDE coding agent (responsible for FPGA programming and investigation) and an Amazon Quick desktop AI (responsible for higher level strategic inquiry, including literature searches, and also LaTeX formatting). At project beginning, the human researcher knew AI engineering but no FPGA programming \footnote{Although by the end he knew considerably more.}. The researcher employed loop engineering principles to insure the agents had clear goals, success metrics, and necessary loop closure tools (including the F2 instance) to allow evaluating their own progress towards the goal~\cite{rostcheck2026loops}.

\paragraph{Implementation status:}
Stage~1 (embedding layer) is validated on physical FPGA (L4) with 100\% accuracy across 500 samples (max error $\leq 0.013$). Stage~2 (encoder layer~0) has achieved timing closure on the VU47P; a deterministic logic-level discrepancy in the attention computation is under investigation. Encoder layers~1--3 and the classifier are validated at L3 (C-simulation). All full-pipeline accuracy claims are based on L3, which has proven equivalent to L4 on validated stages.

\FloatBarrier           
\section{Experimental setup and results}
\label{sec:results}

\subsection{Accuracy results}
\label{sec:accuracy}
Table~\ref{tab:accuracy} compares quantization schemes on the 500-sample test set using C simulation. W16A16 preserves FP32 accuracy (86.54 vs 86.63 F1, within noise) with maximum element error of 0.137 across all encoder stages. W8A8 achieves comparable label-level F1 (85.59). L3 C-simulation matches L2 Python at 100\% label agreement, confirming the HLS implementation is bit-accurate to the quantized reference. The test dataset is small and the task is simply a basic proof-of-concept, as our primary focus was toolchain development, so extensive analysis is precluded.

\begin{table}[ht!]
\caption{Quantization accuracy comparison. FP32 baseline, W8A8, and W16A16 Python ref on 500-sample extended test set. Max element error is measured at the FFN output stage.}
\label{tab:accuracy}
\centering
\small
\begin{tabular}{@{}lcccc@{}}
\toprule
Scheme & Max elem err & Label agr. & F1 & ZIP \\
\midrule
FP32 PyTorch (baseline) & --- & --- & 86.63 & 98.83 \\
W8A8 Python ref & 86.0 & 99.14\% & 85.59 & 99.00 \\
W16A16 Python ref & 0.137 & 99.95\% & 86.54 & 98.83 \\
\midrule
W16A16 L3 HLS C-sim & 0.137 & 100.00\% & 86.73 & 99.10 \\
L4 Phys.\ FPGA (Stg.~1) & 0.137 & 100.00\% & --- & --- \\
\bottomrule
\end{tabular}

\vspace{0.5ex}
{\footnotesize 500-sample test set.}
\end{table}

\subsection{Timing closure characterization}
\label{sec:timing}

\begin{figure}[t]
\centering
\includegraphics[width=\linewidth]{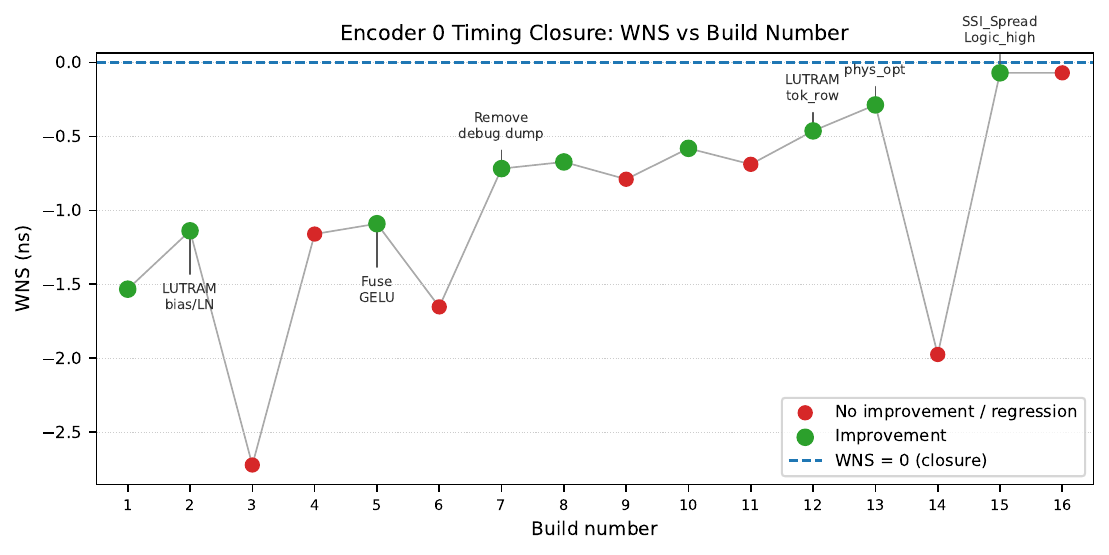}
\caption{Encoder layer~0 timing closure: WNS (ns) vs.\ build number across 16 DCP builds. Green points are improvements; red are regressions. Key fixes are annotated. Non-monotone excursions (e.g., build~3: \texttt{latency=2} applied too broadly; build~6: float sqrt in LayerNorm) illustrate that timing closure on SSI devices requires empirical validation of each change.}
\label{fig:wns}
\end{figure}

Achieving timing closure on the AMD VU47P (3-SLR SSI device) for an ML workload required 21+ DCP builds across two stages. Stage~1 (embedding layer) converged from $-1.533$\,ns to $-0.050$\,ns over 10 builds; Stage~2 (encoder layer~0) converged from $-1.533$\,ns to $+0.001$\,ns over 16 builds. The Worst Negative Slack (WNS) convergence follows a logarithmic decay with non-monotone excursions from failed experiments (Figure~\ref{fig:wns}).

\subsection{Build time and deployment}
\label{sec:deployment}
DCP builds take 38--82 minutes depending on design complexity. Enabling \texttt{phys\_opt} adds 11--41 minutes but improves timing closure by 0.2--0.5\,ns.

\paragraph{Post-bridge patches (required for ML kernels):} (1)~AXI register slice on PCIS path for SLR crossing timing. (2)~VHDL sub-IP copy for HLS-generated float IPs (\texttt{fdiv\_32}, \texttt{fpext}).

\paragraph{Optimization strategies:} Default Vivado directives (not \texttt{Performance\_Explore}) reduce DCP build time by 30--50\%. \texttt{maxThreads=24} saturates the F2's 24 vCPUs. Incremental checkpoints reduce re-build time after minor changes by $\sim$40\%.

\subsection{Performance}
\label{sec:performance}
Because the FPGA build was complete only for the embedding layer and encoder layer~0, we measured that portion across FPGA, CPU, and GPU, using identical inputs. For a short 16-token input (representative of NER), the FPGA is ${\sim}480\times$ slower than CPU and ${\sim}410\times$ slower than GPU (Table~\ref{tab:perf}), almost completely due to the latency of fetching weights from host DRAM.

\begin{table}[h]
\caption{Embedding + encoder layer~0 latency (median, batch~1, seq.\ len.\ 16). GPU fp16 is within noise of fp32 at this size.}
\label{tab:perf}
\centering
\small
\begin{tabular}{@{}lrr@{}}
\toprule
Platform & Latency (ms) & vs.\ FPGA \\
\midrule
FPGA (W16A16) & 344.7 & --- \\
CPU fp32 (c7i.xlarge) & 0.72 & $480\times$ \\
GPU fp32 (g5.xlarge, A10G) & 0.84 & $410\times$ \\
\bottomrule
\end{tabular}
\end{table}

\subsection{Key findings}
\label{sec:findings}
Key findings across both campaigns:

\begin{itemize}
    \item \textbf{Debug writes dominate routing pressure.} Removing the GELU output debug dump alone improved WNS by 0.37\,ns (v2\_enc0g). Debug writes add DDR write calls that increase routing congestion and can place small arrays across SLR boundaries.
    \item \textbf{LUTRAM for small arrays is critical.} Bias and LayerNorm parameter arrays (312--1200 elements $\times$ 4B) mapped to BRAM can cross SLR boundaries under routing pressure, adding 1+\,ns. Forcing LUTRAM (\texttt{\#pragma HLS BIND\_STORAGE impl=lutram}) eliminates this.
    \item \textbf{BRAM latency pragmas are non-monotone.} \texttt{latency=2} on \texttt{input\_i16} (78\,KB INT16 activation buffer) improved WNS from $-0.534$\,ns to $-0.343$\,ns. Applying \texttt{latency=2} to all large arrays (Q/K/V/attn\_out) worsened WNS from $-0.343$\,ns to $-0.649$\,ns as Vivado placed the BRAMs further apart under tighter scheduling constraints. The sweet spot is selective application to the specific array on the critical path.
    \item \textbf{SSI\_SpreadLogic\_high + physopt-loop} produced the largest combined improvement in Stage~2: $-0.286$\,ns $\to$ $+0.001$\,ns. The \texttt{SSI\_SpreadLogic\_high} placement directive spreads logic across SLR boundaries to reduce routing congestion; the physopt-loop (4 iterations with 0.25\,ns over-constraint) then closes the remaining slack.
    \item \textbf{A small ($-0.050$\,ns) timing violation in Stage~1 did not corrupt computation.} The critical path is a pure routing delay (95.5\% routing, 0.165\,ns logic) on the matmul pipeline enable $\to$ K array BRAM write enable. Despite the marginal violation, Stage~1 produces correct results across all 500 samples.
\end{itemize}

\section{Discussion and future work}
\label{sec:discussion}

\subsection{Key results}
\label{sec:key_results}
The FPGA implementation is only partially complete: the embedding layer and encoder layer~0 (1 of 4) are deployed on the hardware. It is, however, sufficient to note several important results:
 \begin{itemize}
    \item The toolchain described in \S\ref{sec:toolchain} effectively enables writing high-level code in Vitis HLS and deploying it to a cloud-based F2 instance.
    \item The 4-level validation process is effective at identifying FPGA-specific problem areas such as broken IP cores and logic manifestation defects, and with sufficient continued effort would facilitate building out the remaining layers.
    \item Performance of our reference TinyBERT implementation is poor vs. a standard GPU implementation, although we expect it could be improved considerably by moving to more standard INT8 quantization now that the toolchain is established and design patterns identified.
    \item Moving weights from DRAM onto the FPGA dominated latency, a cost made worse by the bridge's single DDR port and lack of HBM access (Section~\ref{sec:bridge_limitations}).
    \item Transformers are architecturally optimized for large tensor operations that perform well on GPU hardware acceleration - fundamentally a von Neumann approach. A naive transformer implementation does not fit optimally to the FPGA's multi-layer distributed memory architecture.
 \end{itemize}

 \subsection{Future work}
 \label{sec:future_work}
 Given the above results, pursuing higher performance solutions must contemplate either augmenting or abandoning the bridge, and should consider non-transformer architectures for which distributed memory local to compute elements is more advantageous.

 \paragraph{Augmenting the bridge:} Although we have established a toolchain that allows development of a FPGA design from a high level language (Vitis HLS) to a deployable AFI, this path comes at the cost of greatly limited memory management. One path forward would require enhancing the bridge to support multi-channel DMA access and HBM access. There are no conceptual barriers to doing this. Although it is a large lift, with coding agent assistance, loop engineering principles, and the four-level validation methodology this path appears reachable.

\paragraph{Abandoning the bridge:} Another option is to simply abandon the bridge, which also requires abandoning high-level Vitis HLS and instead implement designs in much lower-level HDL code. Here again the increasing sophistication of coding agents at building low-level code like circuit designs may lower the traditionally high barrier of effort.

\paragraph{Non-transformer architectures and traditional ML:} Work is under way on running non-transformer generative architectures such as Mamba~\cite{mamba} on FPGAs~\cite{mamba_fpga_quant}, but these designs still stream weights from off-chip memory (\S\ref{sec:non_transformer}). The FPGA's highly parallel, distributed-memory fabric is better suited to traditional machine-learning models, which coding agents can now build and deploy with the toolchain we describe. Boosted decision trees compiled to FPGA firmware evaluate in tens of nanoseconds~\cite{summers2020bdt,serhiayenka2025trees}, and online tree learning can run on the incoming data stream itself~\cite{lin2019onlinedt}. Network security is a natural target: traffic is a continuous stream, the market is established, and detection rules must change as attacks evolve. Flow-based intrusion detectors already classify traffic on FPGA SmartNICs at line rate using a few hundred LUTs~\cite{ocuill2026fsnic}, and a coding agent able to regenerate and redeploy such a detector as threats change would make the approach considerably more practical.

\subsection{Reproducibility}
\label{sec:reproducibility}
All HLS source code, validation infrastructure, build scripts, and the complete deployment workflow are available at \url{https://github.com/rostcheck/fpga-transformer}. The four-level validation framework enables any researcher to reproduce our results without FPGA hardware (L1/L2/L3 run locally).

\section{Background and related work}
\label{sec:related}

\subsection{FPGA transformer accelerators}

The FPGA transformer acceleration literature has grown rapidly since FQ-BERT (2021)~\cite{fqbert}, which first demonstrated fully-quantized BERT-base inference on Xilinx ZCU102. Recent work includes ProTEA~\cite{protea} and FAMOUS~\cite{famous} (programmable encoder accelerators on Alveo U55C), ADAPTOR~\cite{adaptor} (runtime-adaptive encoder/decoder), BETA~\cite{beta} (binary transformers on ZCU102), and HLSTransform~\cite{hlstransform} (HLS-based Llama~2 on VU9P). NPE~\cite{npe} introduced an overlay processor approach for NLP on FPGA. The two Waterloo 2025 theses~\cite{waterloo_bekmyrza,waterloo_elayat} represent the closest prior work: integer-only BERT on PYNQ-Z1/KV260 for GLUE benchmarks. Other recent contributions include g-BERT~\cite{gbert} (hardware-aware pruning), Fuketa et~al.~\cite{fuketa2025} (matrix arithmetic-only BERT), BAQET~\cite{baqet} (BRAM-aware quantization), and AutoTrans~\cite{autotrans} (automated accelerator generation). The hls4ml framework~\cite{hls4ml} targets sub-1K parameter models for particle physics applications, a fundamentally different scale from the 14.5M-parameter full transformer addressed here. Table~\ref{tab:related} compares these works with ours.

\begin{table}[t]
\caption{Comparison with prior FPGA transformer accelerators. No prior work targets NER, TinyBERT, AWS cloud FPGAs, or W16A16 quantization.}
\label{tab:related}
\centering
\small
\begin{tabular}{@{}llllll@{}}
\toprule
Paper & Model & FPGA & Quant. & Task & E2E \\
\midrule
FQ-BERT~\cite{fqbert} & BERT-base & ZCU102 & Full quant & GLUE & \checkmark \\
ProTEA~\cite{protea} & Encoder & Alveo U55C & --- & MHA+FFN & Partial \\
FAMOUS~\cite{famous} & Encoder attn. & U55C/U200 & --- & MHA & Kernel \\
HLSTransform~\cite{hlstransform} & Llama~2 & VU9P & Standard & LLM gen & \checkmark \\
hls4ml~\cite{hls4ml} & Small custom & VU13P & QAT & Physics & \checkmark \\
Waterloo~\cite{waterloo_bekmyrza} & BERT & Xilinx & INT8 & GLUE & \checkmark \\
Waterloo~\cite{waterloo_elayat} & BERT & PYNQ-Z1 & Int-only & BERT & \checkmark \\
\textbf{This work} & \textbf{TinyBERT} & \textbf{VU47P (F2)} & \textbf{W16A16} & \textbf{NER} & \checkmark \\
\bottomrule
\end{tabular}
\end{table}

\subsection{Quantization for FPGA transformers}

The FPGA transformer community has generally converged on INT8 or lower to maximize throughput and minimize latency. FQ-BERT uses full quantization, BETA uses binary (1-bit) weights, TerEffic~\cite{tereffic} uses ternary (1.58-bit), FlightLLM~\cite{flightllm} uses INT4/INT8. TATAA~\cite{tataa} uses mixed INT8 + bfloat16, the closest to higher precision. ITA~\cite{ita} demonstrates integer-only softmax at 8-bit. We instead used W16A16 (16-bit fixed-point weights and activations), chosen so that numerical errors introduced by the circuit itself could be separated from quantization noise (Section~\ref{sec:quantization}).

BAQET~\cite{baqet} demonstrates that BRAM utilization is the binding constraint on quantization granularity for stream-based FPGA transformer designs.

\subsection{Cloud FPGA infrastructure}

AWS F1 (2017) and F2 (December 2024)~\cite{aws_f2_blog} provide FPGA instances via the HDK (Hardware Development Kit) framework~\cite{aws_f2_hdk}. Despite eight years of availability, no academic paper has deployed ML inference on any AWS FPGA instance. A 2024 survey of 287 FPGA ML acceleration papers~\cite{yan2024fpga_ml_survey} makes no mention of AWS F2. Montgomerie-Corcoran et~al.~\cite{montgomerie2023fpga_dl} confirm no cloud FPGA ML work through 2023. The University of Toronto explicitly noted the gap for F1~\cite{gao2024multifpga}. HyperAccel's commercial LPU deployment on F2~\cite{park2024lpu} demonstrates that the VU47P is capable of competitive ML inference (1.25\,ms/token on a 1.3B-parameter model), but uses a completely different architecture (a closed-source highly pipelined 4mm ASIC design emulated on FPGA).

\subsection{Non-transformer architectures on FPGA}
\label{sec:non_transformer}

The weight-streaming bottleneck we characterize (\S\ref{sec:bridge_limitations}) follows from the transformer's reliance on large weight matrices that cannot reside on-chip. Architectures with a bounded working set avoid the growing key-value cache. Mamba~\cite{mamba} replaces attention with a linear-time selective recurrence and a fixed-size hidden state. LightMamba~\cite{mamba_fpga_quant}, the first FPGA Mamba accelerator, keeps that state on-chip but still streams weights from off-chip memory, and reaches 7.21 tokens/s on a Versal VCK190. HFRWKV~\cite{hfrwkv} implements the RNN-style RWKV model~\cite{rwkv} fully on-chip for smaller models and reports 32$\times$ the throughput of a GPU on Alveo U50 and U280 cards. To our knowledge, no published FPGA design runs a Mamba model with its weights resident on-chip.

\section{Conclusion}
\label{sec:conclusion}

We presented a complete toolchain for development of ML models on the AWS EC2 F2 cloud FPGA infrastructure, from high-level Vitis HLS code to the deployable AFI image. We introduced a 4-level validation methodology to localize errors to logic, quantization, HLS code, or FPGA circuit implementation. Together these tools allow coding agents to effectively implement designs on a cloud FPGA, testing and debugging until the design is valid. We documented the practical design issues, including various kinds of defects, and their resolutions. For a validation test we implemented the TinyBERT transformer fine-tuned for Named Entity Recognition on a synthetic address data set, implementing the full transformer in C-sim and the embedding layer and the first of four encoder layers on the FPGA hardware. Performance on the FPGA is poor due to latency streaming weights from host DRAM; we discuss the memory access limitations of the community bridge and the directions of future work, including expanding the bridge or switching to lower level HDL, why transformers are poorly suited to FPGA hardware, and which kinds of ML problems the FPGA can best be applied to, particularly now that the toolchain facilitates agentic FPGA development. We hope this work lowers the barrier to cloud FPGA deployment for ML practitioners and opens the AWS F2 platform to the broader ML research community.


\bibliographystyle{unsrt}
\bibliography{references}


\end{document}

%% file: figure_architecture.tex

\begin{figure*}[t]
\centering
\resizebox{\textwidth}{!}{%
\begin{tikzpicture}[
    block/.style={rectangle, rounded corners=2pt, draw=black!60, fill=blue!5,
        minimum height=0.6cm, align=center, font=\footnotesize},
    encblock/.style={rectangle, rounded corners=2pt, draw=black!50, fill=orange!8,
        minimum height=0.5cm, align=center, font=\scriptsize},
    ddr/.style={rectangle, rounded corners=2pt, draw=black!60, fill=yellow!10,
        minimum width=2.4cm, minimum height=0.5cm, align=center, font=\footnotesize},
    shell/.style={rectangle, rounded corners=3pt, draw=black!40, fill=gray!8,
        inner sep=8pt},
    cl/.style={rectangle, rounded corners=3pt, draw=blue!40, fill=blue!3,
        inner sep=8pt},
    arr/.style={-{Stealth[length=3pt]}, thick, black!50},
    darr/.style={-{Stealth[length=2.5pt]}, black!35, densely dashed},
    lbl/.style={font=\scriptsize\color{black!60}},
]

\node[block, minimum width=1.8cm] (host) {Host CPU\\[-1pt]\scriptsize{(Python + fpga\_host)}};

\node[ddr, right=0.8cm of host] (ddr) {DDR4 Buffer\\[-1pt]\scriptsize{weights + act.}};

\node[block, right=0.9cm of ddr, minimum width=1.2cm] (emb) {Embedding\\[-1pt]\scriptsize{+ LayerNorm}};

\node[encblock, right=0.4cm of emb, minimum width=2.4cm] (enc0) {Enc\,0: Attn $\to$ LN $\to$ FFN $\to$ LN};
\node[encblock, below=0.18cm of enc0, minimum width=2.4cm] (enc1) {Enc\,1: Attn $\to$ LN $\to$ FFN $\to$ LN};
\node[encblock, below=0.18cm of enc1, minimum width=2.4cm] (enc2) {Enc\,2: Attn $\to$ LN $\to$ FFN $\to$ LN};
\node[encblock, below=0.18cm of enc2, minimum width=2.4cm] (enc3) {Enc\,3: Attn $\to$ LN $\to$ FFN $\to$ LN};

\node[block, minimum width=1.0cm] (cls) at ($(enc1.east)!0.5!(enc2.east) + (1.5,0)$)
    {Classifier\\[-1pt]\scriptsize{$312 \to 11$}};

\node[block, right=0.4cm of cls, minimum width=1.0cm, fill=green!8] (out)
    {BIO Labels\\[-1pt]\scriptsize{per token}};

\draw[arr] (host) -- (ddr) node[lbl, midway, above] {PCIe};
\draw[arr] (ddr) -- (emb) node[lbl, midway, above] {m\_axi};

\draw[arr] (emb.east) -- ++(0.15,0) |- (enc0.west);

\draw[darr] (enc0.south) -- (enc1.north);
\draw[darr] (enc1.south) -- (enc2.north);
\draw[darr] (enc2.south) -- (enc3.north);

\draw[arr] (enc3.east) -| (cls.south);

\draw[arr] (cls) -- (out);

\coordinate (ddrbelow) at ($(ddr.south) + (0,-1.4)$);
\draw[darr] (ddr.south) -- (ddrbelow) -| node[lbl, pos=0.25, below] {weights} (enc3.south);

\begin{scope}[on background layer]
\node[shell, fit=(host)(ddr),
    label={[font=\scriptsize, anchor=south west, yshift=1pt]north west:AWS Shell (PCIe, DMA)}] {};
\node[cl, fit=(emb)(enc0)(enc1)(enc2)(enc3)(cls)(out),
    label={[font=\scriptsize, anchor=south west, yshift=1pt]north west:Custom Logic (CL)}] (clbox) {};
\end{scope}

\node[lbl, anchor=north, align=center] at (clbox.south)
    {W16A16: INT16 weights $\times$ INT16 activations\quad
     \texttt{ap\_fixed<32,16>} element-wise};

\node[lbl, below=0.12cm of ddr] {28\,MB packed};

\end{tikzpicture}%
}
\caption{System architecture. The host loads tokenized input and W16A16-quantized weights into DDR4 via PCIe. The Custom Logic (CL) kernel reads weights and activations from DDR through a single \texttt{m\_axi} port, processes the TinyBERT pipeline (embedding layer, four encoder layers, classifier), and writes BIO label predictions back to DDR. The full pipeline is validated in C-simulation; only the embedding layer and encoder layer~0 (1 of 4) are currently deployed on the physical FPGA. All inter-stage activations currently round-trip through DDR (dashed arrows); weight caching in BRAM and HBM migration are identified as the primary optimization path (\S\ref{sec:discussion}).}
\label{fig:architecture}
\end{figure*}

%% file: figure1_validation.tex

\begin{figure}[h!]
\centering
\begin{tikzpicture}[
    level/.style={
        rectangle, rounded corners=2pt, draw=black!60, fill=blue!5,
        minimum width=0.82\columnwidth, minimum height=0.7cm,
        text width=0.76\columnwidth, align=center,
        font=\small
    },
    arr/.style={
        -{Stealth[length=4pt]}, thick, black!40
    }
]

\node[level] (l1) {\textbf{L1}: FP32 PyTorch \hfill\textnormal{\footnotesize gold standard}};
\node[level, below=0.5cm of l1] (l2) {\textbf{L2}: Quantized Python (\texttt{ap\_fixed}) \hfill\textnormal{\footnotesize bit-accurate reference}};
\node[level, below=0.5cm of l2] (l3) {\textbf{L3}: HLS C-simulation \hfill\textnormal{\footnotesize Vitis headers, actual kernel code}};
\node[level, below=0.5cm of l3] (l4) {\textbf{L4}: Physical FPGA \hfill\textnormal{\footnotesize deployed AFI, DDR debug dump}};

\draw[arr] (l1) -- (l2);
\draw[arr] (l2) -- (l3);
\draw[arr] (l3) -- (l4);

\end{tikzpicture}
\caption{Four-level validation methodology. Each level is validated against the one above before proceeding. Accuracy metrics are reported in Table~\ref{tab:accuracy}.}
\label{fig:validation}
\end{figure}